# An Interview with Murray Gell-Mann on June 5 2001

Norman Dombey
Department of Physics and Astronomy
University of Sussex
Brighton, BN1 9RH  UK

August 4 2020

**ABSTRACT**

An interview in 2001 with Murray Gell-mann covering his early ideas on parity violation, strangeness, charm, the eightfold way, QCD, electroweak theory and string theory.

**INTRODUCTION**

I had been Murray's student at Caltech and on June 5 2001 we were both in Santa Fe and I thought it would be useful to interview him as I was thinking of writing a new biography of him: George Johnson had just written 'Strange Beauty' and I thought I could do better at explaining the physics involved: this covered his early ideas on parity violation, strangeness, charm, the eightfold way, QCD, electroweak theory and string theory. The biography was never written but the interview may be of interest to theorists and historians of science.

# INTERVIEW

MGM  GJ[1] used the fact that he was allowed to riffle round in the garage to pretend that his book was the authorised biography. It certainly wasn't. It never was.

ND  I'm happy to agree in writing that MGM would see anything that I wrote to comment before publication. We can discuss whether anything I write is authorised.

MGM  Yours can be authorised. I've no problem with that. It would be very helpful if you could help sort the papers. I intend to sell the papers. You're not allowed to give them away in this country to get a tax deduction. They passed a law at the time of LBJ's presidency—the republicans passed a law to get even with him—that you could only claim the value of the paper as a tax deduction. And then the President (Nixon) got round it by back-dating the law. He could have gone to jail for that.

ND  So what would you like me to do in return for going through the papers.

MGM I'd like to sell them to someone who'd give them to a University. I think I can do that. Some billionaire. One of the three or four richest people in the world. I'd like to know what's in there. I'd like to know what's in the boxes. But please be careful of poisonous spiders. There are 3 kinds. The only common kind is the black widow. The female is shiny black. It's small and unlike most spiders it moves slowly. Spiders usually run surprisingly fast. It has underneath, not all that easy to see, a red hour glass. Then there's the violin spider which is a purple violin and the brown recluse. I don't know what that's like exactly but it's quite rare.

ND  I'd better buy a pair of gloves.

ND  I'm most interested in 3 areas of work : essentially the 50s, 60s, 70s. Starting with the 50s, that is the renormalisation group; strangeness and parity. The neutral kaon system. I'd better talk to Francis Low. And to Maurice Levy. That would be the sigma model.

MGM  Well the angle is in that paper with Maurice. We didn't call it an angle. But that's what it is. We called it $e/SQR(1+e^2)$ and $1/SQR(1+e^2)$. But that's the cosine and sine of an angle.

ND Did Nicola[2] know that.

MGM  I think so.

ND  Then there's the work with Goldberger. I don't think that has lasted so well.

---

[1] George Johnson

[2] Nicola Cabibbo

MGM  Well some of it has. We invented the idea (and I carried it further than Murph[3]) of field theory on the mass shell. I did on 54, 55 and my speech in Rochester 56 where I said that you could replace field theory with calculations on the mass shell. I tried for several years to persuade Geoff Chew that this was a good idea. Finally he appropriated it in a speech in 1961and called it S Matrix theory. I said in my speech as a sort of joke that this reminds us of Heisenberg's notion that the S matrix could be guessed instead of calculated. It was part of a programme of extracting exact results for strong coupling from field theory. I think that has lasted. The whole programme of string theory and M-theory is on the shell.  In fact with Murph I did low energy theorems and dispersion theory. With Francis I did the renormalisation group. The one thing that has not lasted especially well is the formalism we developed for scattering quanta by quanta so that could one have a single equation connecting all the amplitudes and we wrote it as a sort of  field theory. But it was a field theory of Feynman diagrams rather than a normal field theory. It was part of a programme I was following of extracting exact results for strong coupling. With Murph also we did the crossing relations. Geoff then rechristened them the substitution relations.

ND  One thing I could do, but I've been advised against it, is just to read all the papers and then describe them for physicists. But I don't want to do that. So I have to choose what seems to be the most important.

MGM  One of the most important developments in the last half-century has been the interplay between field theory and the theory  on  the mass shell. I always thought that they were 2 different ways of saying the same things. Geoff Chew and Landau tried to say that field theory was rubbish. It turned out that what they were objecting to was not field theory, but non-Yang-Mills field theory. As soon as you have Y-M with limitations on the particles you are introducing you have an asymptotically free field theory and all the things that they were worried about go away. This was also what leads to superstring theory. The way I view the transition for a unified theory. It came from the work on the mass shell; the work of Geoff and others who tried to solve the pion-pion system by exchanging a rho, and my complaint that you needed an infinite number of baryons and an infinite number of mesons in there. You could approximate them by stable particles initially and then  perturb the system by allowing the particles to disintegrate. That was called the duality idea and was written down by my post-docs Dolen, Horn and Schmidt. Whereupon Veneziano produced the Veneziano model (it is wrong of course) but that is string theory. That is the basis of all this and that is on the mass shell. Now is that the same as field theory. It has still not been decided. It is not known whether you can do M-theory with fields.

ND  But QCD you cannot do on shell.

MGM  That's another exciting thing. The whole theory is on shell but none of the particles show up. This is the problem of confinement. This remains  a very live issue over 50 years.

ND  Let's go back to the renormalisation group. GM-Low was 20 years before its time.

MGM  But 2 other people did it. Petermann and Stueckelberg did it too.

---

[3] Murph  Goldberger

ND  And Bogoliubov and Shirkhov?

MGM  Bogoliubov stole it. Just stole it. He and Shirkov wrote a paper which was an exact copy of ours. Even the equation numbers and symbols were the same. But there's a reference to us. The reference doesn't say this paper is entirely stolen from GM and Low; it says GM and Low tried something along these lines but were wrong. And what was wrong was that we used the wrong gauge. How do you use the wrong gauge in a gauge-invariant theory.

ND  Did you know Stueckelberg. I once gave a seminar at CERN and Stueckelberg sat in the front row with his dog.

MGM  Yes I knew his dog. He was crazy of course. He kept having periods of insanity and tried to resign. Petermann was also crazy. I was also friendly with Petermann. I invited him to Caltech but he never turned up. At CERN he worked only at night. It was like Fritsch's idea of a night shift and a day shift in the theory division. He was the night shift.

ND  Well who should I talk to about this period. Low, Levy and Murph I guess. Let's turn to parity. According to GJ, you were given an graduate problem on this.

MGM  I wrote this in one of my reminiscences. When I was a graduate student, Herman Feshbach assigned a problem in his class to prove by a transformation of coordinates that parity is conserved. I thought about it for a time and realised it was nonsense. You can't prove anything by transformation of coordinates. You need a theory which is invariant under transformation of coordinates. If the theory is invariant then you have parity conservation. If the theory is invariant under a parity transformation then parity is conserved. If it is not, then parity is not conserved.

ND  You said that.

MGM  Yes. Of course in QED parity is conserved. But in some other theory it need not be. I ran across this 1+gamma5. I brought it to Feynman and said this explains why the neutrino is massless.

ND  He didn't know that.

MGM  Well I suppose he did. I like to collaborate. He likes to show how smart he was.So he said I don't know about that. So I didn't do anything about it. But then at the Rochester meeting when we had the tau-theta puzzle, Feynman's roommate Marti Block proposed that parity need not be conserved.

ND  Yes I know that.

MGM  But you don't know my story. Feynman came to me and said "my roommate Marti Block suggested that parity need not be conserved. I said in the weak interaction we have no idea whether parity is conserved. And you remember last year I showed you this 1+gamma5. He said yeh I remember. So you agree that there is no experimental proof that parity is conserved. "In the weak interaction" I said; in the weak interaction we have no idea whether parity is conserved.

ND  And this is before Lee/Yang.

MGM  Jeremy Bernstein wrote all this up in the New Yorker but he doesn't know about me. Yang was the chairman of the session the next day. Feynman got up and said "my roommate Marti Block has suggested that maybe parity is not conserved. BUT HE DIDN'T SAY IN THE WEAK INTERACTION.

ND  Didn't he realise that?

MGM  I suppose he did. Sometimes he had an extremely precise vision of things. Sometimes he didn't. He didn't say that the chairman should ask me. At that time I was preoccupied with this other theory; that the tau and theta were a doublet. And therefore the lambda would have to be a doublet; the sigma would have to be a doublet, and the xi would be a singlet or a triplet under some new symmetry. So I didn't pursue this. And Feynman didn't pursue it either. He was very angry with himself for nor pursuing it. Yang and Lee revisited it in the summer. I suspect it was Lee as Yang loved parity. What Yang said at the meeting was Lee and I have looked at this and it doesn't lead anywhere. So then I went to Moscow in May 1956. In Moscow I gave a speech on the tau-theta puzzle which was so well attended that about half the people couldn't get in. So I had to give it a second time. Everybody was there. Landau was there; Kapitsa was there. I listed 3 explanations. One was the doublet explanation of mine, which was subsequently published by Lee and Yang. The second explanation was Marti Block's that in the weak interaction parity was violated. And they were furious. People got up and said you are violating Lorentz invariance. I said this is an empirical matter.

ND  CPT theorem was known wasn't it in 1956? Why should they be so uptight about parity?

MGM  It was known. But they thought like Feshbach. I told them it was an empirical matter. I told them that it was simply not known experimentally in weak interactions whether parity was conserved. Both MGM and Feynman had looked and couldn't find any indications one way or the other. I don't know why I didn't say "test it by looking at the polarisation of the electron". I could have said that but I didn't. The third explanation was Marshak's that it has spin two and could decay either way. But then Landau beat his young people like Okun for suggesting parity was not conserved and finally he adopted it himself. But I didn't make much contribution to that: I could have but I didn't.

ND  Is there anyone else I should talk to about the 50s?

MGM  Okun.

ND  Who's that? [turning to photograph in study]

MGM  Zeldovich, in Chicago

ND   Zeldovich was in Chicago?

MGM  Yes, before he died. This was 1988. I told him that I thought that he was a greater physicist than Landau.

ND  That must have pleased him.

MGM  It sure did. But he said "No. Landau was a master"

ND  Sasha Polyakov doesn't have such a high opinion of Landau.

MGM   Landau was a smart fellow. But he was very bigoted. Zeldovich was very reasonable. But he co-invented the hydrogen bomb with Sakharov. Finally Gorbachev let him out. But neither lived to see the end of the Soviet Union. The Landau picture was taken when I was somewhat younger than that.(1956). I don't look too different except that I have dyed my hair.

ND  You know that when I left Caltech I had hoped to work with Landau.

MGM  After the accident. You know they told me that if I had gone to see him that he would have recognised me and spoken to me in English. But 15 months later he would have forgotten all about it.

ND  Do you really not forget.

MGM  Now. I forget names all the time. Eventually I remember them. But it may take all day. But I used to have a very good memory.

ND  But you weren't a mnemonist. Someone who doesn't forget anything.

MGM  No I don't believe that is possible. Landau was also reputed to have made a joke after the accident. He was asked whether he would be able to do physics again. He replied "Maybe. But I'll never be able to do physics like Landau. Maybe I'll be able to do physics like Zeldovich"

ND  So I should talk to Okun about your time in Moscow?

MGM  If you like. My story is true. But there are lots of other narratives. People have other recollections.

ND Telegdi?

MGM  He was certainly there. But he is very odd. God knows what he'll say. He says some bizarre things. He put up a little plaque at the desk in his office where I worked out the strangeness stuff after about a year. It took me a year to write it up.  I was about to be drafted and used Val's office as it was air-conditioned for the equipment. Equipment needed air-conditioning but theorists didn't. You were talking about Levy in Paris. The main thing I thought about all the time in 1958/59 was Yang-Mills theory and how it could be generalised. Shelley[4] came to visit me in Paris. I liked his ideas. The day I came back was the Rochester conference 1960 where I spoke about Shelley's ideas. He hadn't spoken about them in public. I acknowledged that they were Shelley's and spoke about them in a very clear manner. He had never spoken about them so clearly. I spoke about them as $SU(2) \times U(1)$ and so on with certain term breaking it. In the meantime I was thinking about how to generalise Yang-Mills. But I had forgotten about Lie groups. I have written about this in the reminiscences. I generalised to $SU(2) \times SU(2) \times U(1) \times U(1) \times U(1)$ : up to 7 dimensions in the theory. Not in the representations but the theory. I knew that there was nothing else with a single charge. The person I was having lunch with every day was this famous mathematician who was an expert in Lie Groups. But I didn't ask him. Of course if I had it wouldn't have helped as the language was all wrong. When I got to Caltech there was this Assistant Professor named Block who was simple-minded enough to look at my commutation rules. He said that this was just representing Lie Groups with unitary representations. You take a Hermitian operator which generates unitary representations. This was all classified by Elie Cartan. The one you are interested in is the unitary group with three. Shelley was away. He was in Boston so I didn't discuss it with him. I wrote it down on January 12 1961. What I want in that garage are the 3 pages. They are on yellow paper written in pencil. I sent them at least one page Possibly three. And on January 20 I published a Caltech report. But I didn't realise it wasn't a publication. It was treated as a preprint.

ND  A lawyer would have said that you were right. LANL took legal advice over their archive.

MGM  But that doesn't mean that the scientific community agrees that is how you establish priority. Newton thought that the way to establish priority is to write one or two words in Latin describing the work and hide them under a rock. Then if someone else has the idea you say look under the rock. Leibniz published. It was decided that priority goes to the first person who publishes and it stayed that way until Ginsparg came along. And my colleagues told me that it was not a publication. Anyway those pages. They were abstracted from my garage. And they were displayed in Washington and then returned to me from the Smithsonian. I desperately need those pages.

ND  When were they returned. When are we talking about? 20 years ago?

MGM No. I was a Regent from 1974 to 1988. So it would have been something like 1980. If it was then, Margaret was dying. I was very confused. But I certainly didn't throw them away.

ND  Do you have the date of that lecture you gave?

MGM  It would have been January 1961.

ND  But the precise date.

MGM  I don't know. You would have look at Caltech's records. Was it a huge lecture?

---

[4] Shelley Glashow

ND  It was enormous.

MGM  We were still using 210 Bridge.

ND  I don't think it was there.

MGM  It would have been in the Humanities Building. I don't remember the date.

ND  I would have thought that it was before the preprint came out.

MGM  The preprint is dated January 20 1961. Eight days after the January 12 synthesis that I wrote down in my own notes.

ND  I reckon that the talk must have come out between those two dates.

MGM  Shelley wasn't there when I talked with Block. But he came back after I wrote the synthesis. I worked on it over the Christmas vacation. Shelley would have been at the talk. And Sid Coleman

ND  No he wasn't.

MGM  Sid was a graduate student. He worked immediately with Shelley on R-invariance, which was a symmetry between the nucleon and the xi. They concluded that there wasn't much symmetry between the nucleon and the xi. It was a sort of parity for SU(3).

ND  I think Sid was at Harvard during this period.

MGM  Why should he be at Harvard? He was a graduate student at Caltech.

ND  I think Shelley arranged something for him. I think he was at Harvard at that time.

MGM  Neither of them had any connection with Harvard.

ND  Maybe it was 1961-62. Shelley and Sidney may remember.

MGM  They worked together on R-invariance. Then they did the em mass differences. I think it was Squid's dissertation, wasn't it. I have all the dissertations in the garage.

ND  The whole business of quarks. Or is it quorks?

MGM  In Joyce it is certainly quarks.

ND  I guess I should talk to George again. And to Dick Dalitz while he's still with us. Dick's in pretty bad shape.

MGM  I'm sorry to hear that. Dick has spent so many years at the Wem. I saw him once in Washington. By chance. He was very embarrassed. He said he was going to the National Archives. Of course I was very interested. All these names ending in "itz".

ND  Oh incidentally. Harold Morowitz. Where is he at the moment.

MGM  He is at home in New Haven. But he works in Virginia. At the Krasnow Institute at George Mason Institute. He came from Poughkeepsie in New York. But his father or his grandfather would have been born abroad. They would have spoken Russian. They owned a newspaper delivery service. In 1946 Harold and I got a ride from Poughkeepsie to the Berkshires. Sheffield Mass was the ultimate delivery point for the newspaper delivery service. From there we hitchhiked to Springfield and then on to Boston. I had never seen those places before. I had never seen Mass before and it was very exciting. I had been to Connecticut.

ND  That was before you went to Yale.

MGM  No after. We were lab partners. So I have known him for 56 years. So how did they write that in England.

ND  Morovitch. Well now period three.

MGM  Wait a minute we haven't talked about period two.

ND  OK you say something. The point is I was around during period two so I think I have a good idea of what happened.

MGM  Quarks, QCD, the revival of PCAC. We realised that certain symmetries (the axial symmetries) were realised through near zero mass objects rather than near current conservation. I forgot that for a while and wrote some notes where I tried to treat the axial symmetries as degeneracy symmetries instead of near zero mass particles. That was important. Even that stuff with Bruno Renner.

ND  It was important but not that important.

MGM  It was important because it is a question of how the symmetries were violated by the expectation values of qbar q in the vacuum. And that's really important.

ND  Yes it leads to anomalies.

MGM  That's right. The anomalies were very important. I didn't do much on them but my guests did especially John Ellis. I should have paid more attention to what he taught us. He called them POT (partially ordered trace). I wouldn't have worried so much about the trace with QCD in the limit of zero quark mass. I shied away from QCD at first. I thought it up with Harold: I then had all these qualifications: it's all written up in the reminiscences paper "Quarks, Colour and QCD" but one thing I worried about was the violation of this symmetry. But I didn't have to worry so much because the trace has an anomaly. So you don't have to worry that in the limit of zero quark mass the trace seems to go away. But it doesn't go away because there's an anomaly. But I didn't remember about the anomalies.

ND  The problem you set me I did not do: you wanted me to derive the Adler-Weissberger relation. I nearly did it the year before Adler in Geneva after I left Caltech but I couldn't do it because the equal time commutation relations were not covariant and I did not know about going to the infinite momentum limit because it hadn't yet been invented.

MGM That was another thing we did in the 1960s. I could have done all this crudon stuff also because what I did was to ignore the distribution in longitudinal momentum fraction so I just looked at the transverse situation and assumed that you would just have to ignore pairs: then Roger Dashen showed that the only solution was free particles.  But then if I had included the longitudinal momentum fraction as a variable then that would have been the whole story.

ND Anyway it is a interesting decade and clearly one which has to be pretty central to the story.

MGM  Also the same decade was the revival of the weak interaction story.  The revival of Shelley's theory which I presented in Rochester in 1960 and then this idea. I told you that when I returned to the country from East Africa in the fall of 1960 the first thing I did was to go to Rochester.  Directly from New York.  From Rochester I delivered the speech about Shelley.  Now in the 60s it was revived and in the later 60s along with the sigma model idea of the Higgs boson idea and someone showed at the end of the decade that the theory was renormalisable.

ND Shelley was actually very unfortunate because you can actually derive from his theory the mass relation between the Z and the W which was the Weinberg's contribution.

MGM  I did not call it W: I called it X+, X – (uxl) and then Lee and Yang wrote it up without referring to us.  They did some very good work because what they suggested was that you could approach it through the neutrino beam: that was a brilliant suggestion which never occurred to Feynman or to me that you could have neutrino beams and that the weak interactions would keep rising in strength up to a certain point which was the mass of the uxl and that therefore you would get large cross sections for neutrinos.  We just did not think along those lines.  We suggested the red and blue neutrinos.  Feynman repudiated them by the way: he constantly repudiated our ideas and when Bludman talked about them a year or two later at Gatlinburg he laughed at Bludman but I didn't, I liked the red and blue neutrinos.  I didn't see any way of checking the theory but Lee and Yang did.

ND That's interesting because I worked on neutrino scattering and think that it was obvious that the cross section increased with energy.

MGM  Of course it was obvious it didn't occur to me that gave you a clue as to how to, no we worked out a formula for the cross section but I did not see that that made the cross section big enough so that you could have a neutrino beam and you could have experiments.

ND But that Lee and Yang paper was very early. Was it in the 1960s?

MGM 1959, but I had been working on that for years. The intermediate boson. Again Feynman did not want to write up that work that we did on mu to e + gamma without red and blue neutrinos. We worked at it extensively and even did it with the correct theory, the present day theory that was the only one that gave a finite result. Only if you had the Yang-Mills kind of coupling did you get a finite result from mu to e + gamma assuming the neutrinos were identical. And that finite result was wrong: mu didn't go into in e + gamma that fast so we concluded there were red and blue neutrinos, and Feynman repudiated this.

ND This was without W.

MGM No this was all with the X. We did it all with the Xs. This was in 57.

ND You are right if you take the mass of the intermediate boson to infinity then it diverges.

MGM We had everything perfect and we concluded that there were red and blue neutrinos and I wrote it up actually but Feynman refused to sign it so I got very depressed and discouraged. Then this guy at Columbia Feinberg he wrote up somewhat the same thing so he is given the credit. Then Lee and Yang suggested the two kinds of neutrinos but we already had them … however, I got discouraged. There are a whole lot of things he would not sign, the non linear sigma model, I put his name on that, but he said take it off. I don't believe in square roots of an operator? I shouldn't have listened to him. I should have listened when he was constructive and should not have paid any attention when he was destructive. When he was constructive he was great, but of course when he was constructive it came out that he had done whatever it was, not we, but if he was destructive, then I had done it.

ND But that so often happens.

MGM It was my fault for paying attention.

ND Polyakov invented the Higgs but Gribov wouldn't accept it for publication. People have set ideas.

MGM Feynman had very good ideas about this. He just repudiated them.

ND We all have good ideas and not so good ideas.

MGM It was a strange period a very strange period.

ND It was an intensely productive period.

MGM All the things I had, all the things I missed and I did not do anything with.

ND Well you can't do everything. You got credit for plenty. OK, QCD I should talk to Fritsch and the third area of the 70s would be Strings.

MGM  Fritsch and Minkowski worked  as well as the Harvard people on the unified theory: unified Yang-Mills theory. I would certainly not call that the grand unified theory because first of all the previous theory  did not unify the weak and electromagnetic interactions: it just glued them together.  This is not grand unified, it is also not grand because it does not include gravity.  So I call it unified Yang Mills theory.  Which I think is a good name.  Fritsch and Minkowski were working on it too. I do not know why they did not publish it in time, or did not get credit,  or what the story is but it was an interesting idea  and at Caltech we were thinking about that too. The curves would cross at a very high energy.   I felt that Shelley should share the Swedish prize that they wanted to give to  Weinberg and Salam and I felt that Ward should share it as well, that would be four people.  They don't give it to four people they gave it to three, for some traditional reason that I don't know about.

ND Salam didn't do anything which wasn't in collaboration with Ward.

MGM  Exactly. That is what I told them.  I also told them that Shelley had done more interesting stuff than either Salam or Weinberg.

ND So did Ward for that matter.

MGM  Probably.  I said I was happy with Salam getting an award he has done some very good things.  But with this particular problem he hasn't done anything that Ward hadn't done: I let them know that, and they gave Shelley a share of the prize, but I am not glad of  the use he's made of it since. He's used the prize to make fun of string theory and it is ludicrous.

ND Well people can have their own ideas.

MGM  But it is absurd: he says there is no way to verify it because the unification occurs at a very high energy.  Now people who live in wooden houses shouldn't throw termites.  He and his friends are the guys who introduced the unified Yang-Mills theory which has a very high energy.

ND  But not such a high energy.

MGM  Approximately the same. And there are plenty of ways to test the theory.  That should be tested and it has been tested through its prediction of general relativity.  Anyway, I am giving you plenty of my points of view.

ND  And on strings?

MGM  I told you the story from my point of view and there are many different stories.  From my point of view it was that  I loved the bootstrap idea. I think the bootstrap idea is still a very useful idea.  Still correct and still… find out exactly how to say it right.  It is very different from the Gross person who says things differently. I think the bootstrap is a very brilliant idea that it underlies string theory and that it is the ancestor of string theory intellectually. First of all there was the work on the shell which Chew finally accepted in 1961 after various Russians had formulated the problem in this way. Chew and his collaborators beat to death one or two statements in their comments on how the bootstrap would work. I suggested that it be with an infinite number of baryons and an infinite number of mesons and you work outwards from an approximation in which they are stable: in this case the Veneziano model. Well it was written done by Dolen, Horn and Schmidt in the duality paper which led to the Veneziano Model. The Veneziano model was string theory, which was shown by all the nice people who sat in the next office at CERN: Peter Goddard, Claudio Rebbi, 4 or 5 people. They showed it was a string theory although a bizarre string theory. Then in 1971 John Schwartz and André Neveu invented the superstring theory which did not have any of these problems but initially it was not clear that it had no problems. But over the next three or four years it became clear that there were no problems. But they did not know what the group was; they thought it was the unitary group but it turned out to be the orthogonal group: accompanying $SO(32)$ was an $E(8) \times E(8)$. I kept begging Schwartz to use the $E(8)$: I said it has got to be $E(8)$ IT HAS TO BE $E(8)$.

ND $E(8)$ is an exceptional group. But why had it to be $E(8)$.

MBM I showed that other groups had led to the wrong kind of symmetries but the other thing is that it is a group where everything is in the same representation. It is exactly what you need. I insisted that it had to be $E(8)$. I did not know that it was $E(8)$ times $E(8)$ of course but they showed that when they worked out $SO(32)$. Then the Princeton string quartet came in. Now of course they've all been shown to be part of the same theory. But the essential thing is how to state the bootstrap because the bootstrap properly stated is the underlying principle of the theory. The vexed question of how field theory and the shell are related is still with us and is still very important.

ND Now the string theory work you encouraged.

MGM I did not do it myself but I encouraged it very strongly. I immediately hired John Schwartz first year and I hired Pierre Ramond as soon as I heard of him. I established this nature reserve for endangered string theorists and then we brought over Scherk, then he and John showed that the theory could be revamped as a theory of all particles, not just strong interactions.

ND Mike Green?

MGM Yes Mike Green

ND He came to Caltech also.

MGM He came to Caltech along with John Schwartz and worked out their actions at infinities and all that.

ND OK I can talk to him.

MGM There were other collaborators as well.

ND That is not everything but the main substance of the physics.

MGM  In the Erice series of reminiscences  Gross gives an entirely different picture of what happened.  He apparently had some very good ideas before QCD  by a different set of reasons.  What annoyed me about him was that he was the Chairman of the meeting in Chicago where I presented the speech about what was later called QCD but I repudiated it unfortunately in the write up where  I was very vague, but in the talk I talked about exactly what was QCD.

ND  What year was that?

MGM  1972

ND As early as that.

MGM  In 1972. I came back from Europe and just as I had gone to Rochester  in 1960 I went directly to Chicago in 72 and I spoke at the meeting and what I described  was QCD but in the write up I became very vague and retreated into formalities. The reasons I explained  in the reminiscence article was that I was worried about the zero trace in the limit of  zero quark mass and I was worried about some other things as well.  They were all very important things that I was worried about like the cancellation of the axial current; but they were all resolved using the anomalies of course  but  I hadn't paid enough attention to them, I knew about them. Caltech had had lots of talks about them from my guests but they hadn't registered enough.

ND The other areas?

MGM  You have the colour business in 1971.

ND You distinguish that from QCD.

MGM Yes, because that was a very clear cut thing.  Harold Fritsch and Bill Bardeen and I wrote that up in the Fall of 71, and what we said was that parastatistics with no para particle was just like colour, with no colour particle.  Coloured particles were confined and  would not appear on the shell, and then we said that colour could be very important and that in particular  it would restore the pi 0 to gamma gamma amplitude which was otherwise  too small by a factor of 3.  I had this bet with Heisenberg, which he lost and never paid.

ND I  did not remember that you did that.

MGM  It was very annoying because people like Sid  Drell and so on laughed at it; they said it was ugly, stupid and they contrasted  coloured quarks with what they called Gell-Mann and Zweig quarks.  Which were the ones without colour and what was particularly annoying was that it had my name attached to it.

ND For some reason I thought of colour as obvious with the quarks.

MGM  It was invented earlier by Nambu.  Nambu and Han had this tripling of the particles in order to get integral charges and they had therefore  a coupling of the electromagnetic field to the excitations of the theory which meant they had to be real.

Nambu then invented what amounted to QCD, or the beginning of QCD, in the 1966 paper; in fact if I had read it I would immediately have gone through the whole thing. I was so ashamed of not contributing to Weisskopf's festschrift that I wouldn't read the articles so I did not read Nambu's article. Now of course he had it coupled with all these wrong ideas of the integral charge, of the electromagnetic field being able to excite the degree of freedom. I would not have done that as I would have done it correctly. If I had read Nambu's paper it would have set me ahead by several years. Although I really have to give him a huge amount of credit even though he did not follow it up when he had it wrong. But what Harold and Bill and I did was to put in colour correctly and it was very good: we were very happy about that. We should have just gone on to QCD. There was another thing that was bothering me about QCD, that was I was convinced that the correct theory was a string theory. Because these guys in the next office at CERN Peter Goddard and all these others, Rebbi and David Olive too I guess, they were showing the equivalence of all these models I liked so much and string theory so I thought the true theory must be string theory. I didn't realise that it was at another level that we would have strings. This level we would have ordinary field theory. My whole life was like that and I think many people's lives are like that. If you generalised my errors and the things that I got wrong and the things that I didn't follow up properly and the things that I saw and I did not believe in and the things I saw and did not write up. Almost always there was some error of level. That is I knew that a certain thing was right and felt that it was right and it was contradictory to what I was doing and I could not get used to the idea that some things you have to answer late: you just put them off, but you answer some of the things now and ….

ND I thought that was what you actually did; that is when I was at Caltech you had all that stuff about vector particles and gauge particles and you said that they have a mass. We don't know where the mass comes from but we know it breaks the symmetry.

MGM I said it came from some soft mechanism yet to be discovered. But I had discovered it, it was the sigma model.

ND The sigma model wasn't as clearly described as the Higgs model.

MGM It's the same thing.

ND I know it's the same thing but it nevertheless it wasn't a mass generating mechanism.

MGM I didn't know it was mass generating that what I just said. I didn't know it was generating masses. But it was a mechanism that I already worked with. I advertised in a preprint for a soft mass generation mechanism and I had invented the mechanism.

ND Peter wasn't trying to generate mass either.

MGM Peter Higgs, I don't know what he was doing. I never met him.

ND Anyway he's a very nice guy.

MGM  I understand he was a wonderful guy but I just never met him.

ND He was trying to find an exception to the Goldstone theorem.

MGM  You mean when you get a zero mass particle by breaking a symmetry if you don't have degeneracy. One case you have degeneracy; the second case you have zero mass bosons, the third case you have zero mass bosons that get eaten by zero mass vector bosons that give massive vector bosons.

ND That is what he was trying to do and he did.

MGM  He did, he found the third case.  It was very clever and something I didn't pick up on.  But I did advertise for that soft mass mechanism.

ND  What I was saying was that the violation of gauge symmetry bothered you but you were were pragmatic enough to say "what the hell" and go on.

MGM  When I did that that was when things worked.  When I did not that's when I got all screwed up about it.  Failed to publish important things, failed to see important things because I was preoccupied with a real problem, a very serious difficulty, but I didn't acknowledge that that difficulty would be resolved at a higher level.  I think that that's a thing worth telling people. I suspect that's a really general principle which may apply in many fields and different aspects of life in a huge variety of cases. Only if you can distinguish between the things you can do something about now and things that you have to put off and do something about later can you make progress.  It's so important. Otherwise you get paralysed.

ND Changing the subject. I don't want to talk about complexity. I'll ask Geoff[5] about that and in any case the focus of the book is you and elementary particles.

MGM  You don't want to talk about conservation and nature.

ND  May be.

MGM  It would be amusing to have a brief discussion.

ND  Let's talk about that, I said I would talk on the same, you make some interesting points you are interested in birds, and nature and classification and taxonomy.That to me is very strange for a physicist to be brought  up that way.  That is a different way of doing science.

MGM  However it is a general thing in science that structure should come before mechanism and in most fields they are very far apart, you learn something about the structure and a century later you learn something about the mechanism, like genetics, DNA and all these other things that make sense of evolution. Mutations were discovered in 1900 but no one knew what they really were until so much later in the 20$^{th}$ century.  But in particle physics the magic of gauge theories  means that when we understand the symmetry principles we have the dynamics free.  That's amazing.

---

[5] Geoffrey West

ND  Are you saying that that was what was driving you?

MGM  That was what made the whole thing work.

ND  But you only learned that after the event.

MGM  I was brought up to think about classification, taxonomy, nature and I said many times when people have asked me to write down a biographical note that I applied even in physics the style of someone interested in natural history.  You know that when Howard Gardner listed many types of intelligence he showed them to Hans Meyer and Hans Meyer said you left out natural history intelligence and he put it in.  So most physicists don't have natural history intelligence but I do.  However what made it so magical, made it possible to do so much with that attitude (the attitude might have led nowhere)  was gauge theory.

ND  Was it gauge theory or was it group theory?

MGM Gauge theory. It meant that when you had the symmetries which are a structural feature…

ND I remember you saying when I was at Caltech, that in mathematics you were most comfortable with algebra.

MGM I just love algebra.

ND That was when you were giving those talks in the early days it was the patterns you were getting.

MGM  It was the natural history of the particles. Algebra is discrete and algebra is descriptive…

ND But algebra is not necessary gauge theory. That's what I am saying.

MGM  I know but it's gauge theory.

ND  Now we know that gauge theory works.

MGM  The magic of gauge theory means that when you have these simple natural history features that we call symmetries they GIVE YOU THE DYNAMICS FREE. That's what so great. It's true with Einstein and general relativity too. So in other fields it is often very important to look at the structure and put off the dynamics till later.  But in our field there is not this gap, this century of waiting.  This is just the point I make from time to time. I have just made it on television shows a few days ago.

ND  Lets change the subject again.

MGM  You don't mind that I tell you all these enthusiasms.

ND  No it's what I came for. Or at least one of the things. I'd like to see your papers.

MGM  Yes there's also the garage.  A volunteer coming 6000 miles to clean my garage.

ND  The ABM Treaty.

MGM  I did not have anything much to do with the treaty. I did play a role in the arguments. It should be called active anti-missile defence of the cities.  Or of populated areas.  Any of those.

MGM  In the summer of 1962 we had a study in Berkeley (a JASON study)  JASON was very new then but I was one of the founders in 1959-60 (but I was in Europe that year). When I got back I found that I had all sorts of clearance problems which were finally settled  a few months later and so although I was one of the four or five people who started it, I didn't really participate until early in 1961. In the summer of 1961 I was in Berkeley doing particle physics but in the summer of 1962 we had a JASON summer study and I chose always to be interested in strategy. I was interested in the notion of active anti-ballistic missile defence of cities and I thought that would be extremely destabilising in many different ways which you know and so I worked on that and suggested that it would be a very good idea both for the US and the USSR to hold back. I had a very long discussion of it with all the usual things: the issues aren't very different from today. They are somewhat different but not very different: how it probably wouldn't work very well; how it could be circumvented in all sorts of different ways; how it would lead to increases in offensive forces;  but also that it would lead to crisis instability and passive defence likewise. Sheltering populations would give an indication of the intention to strike first.

ND  What do you mean by passive defence

MGM  Shelters. I suggested therefore that it was a rotten idea and both sides should refrain from it and that they might  have some sort of informal understanding. I despaired of having an actual treaty which would be difficult to ratify. When I gave the speech one or two people said that they agreed with what I was saying but that maybe it could be a treaty. I said that I didn't object to a treaty but that it could be done if necessary without a treaty.

ND  This was summer of 1962 even before the test ban treaty?

MGM. Yes. I didn't write this up formally but I gave some speeches in Washington and briefed people high in the Defense Dept and so on but they had already thought of all this but hadn't done anything about it. Then in January 1964 I was in a Pugwash meeting in India, the only one I've ever gone to, where I presented this kind of stuff in an unclassified way to an international group along with Jack Ruina (who had been the head of ARPA) and Carl Kaysen (who had just stopped being the deputy national security advisor to the President) and was later going to be the head of the Institute for Advanced Study. So Carl, Jack and I presented this paper which I described somewhere in one of my reminiscences. We first presented it to the Soviet delegation and they received it very badly and barely could keep from laughing and said that if we subscribe to this we'll never get another vote in the next election. "If the Soviet

Government adopts this position it will not receive a single vote at the next election". Then later in 1964 ....

ND  This was after the Cuba crisis when both sides were anxious to have some sort of common understanding?

MGM  Yes. But the Russians were gung-ho for ABM. Doty and Kistiakowski and Carl went to Moscow to talk about it. The guy who had led their delegation and had laughed the loudest at our proposal in January was Milionshchikov. Physicist, Vice President of the Soviet Academy, a big-shot. They got it to various people including even Kirillin, on the Poliburo and helped to change their minds. There was a funny incident involving this little swine… I've forgotten his name. He was the Secretary of the Academy of Sciences and obviously a KGB type person. Polichenko. He greeted the people and accompanied them to their hotel and all that. George was accompanied by his wife Elena because he was a Russian by birth. He received the royal treatment and beautiful room and all that. The other guys were rooming with each other as they were not Russian. They got some relatively crummy room in the hotel. So they said to each other in their room "This isn't very good compared with what they gave George". They went downstairs and Polichenko said "I hear you don't like your room." They talked to various people and presented the same point of view which Jack and Carl and I had presented in January in India. Next in the summer same year I was back in Russia at the International Particle Physics meeting in Dubna. I had just heard Cronin and Fitch talking about their experiment and was sitting at the very back of the cafeteria chewing on some food and writing on a pad of paper trying to think what it would mean that CP was violated. I was playing with the idea that it was some spontaneous thing with an expected value and why would it be that when I noticed out of the corner of my eye some stocky man who was walking around the cafeteria from table to table looking around. I didn't pay much attention but then he came right up to my table and I noticed that it was Milionshchikov. What the hell was he doing at this particle physics meeting? He didn't know anything about particle physics. He came and sat down at my table—the two of us were the only ones at my table—and he said "you remember me?" I said "I remember you very well Akademik Milionshchikov. He said "You remember we were at a meeting in India together?" I said that I did. He said "You remember you presented a paper with two other people on missile defence? I said that I did . You remember what I said about it?" I said "Yes. You said it was crazy". He said "Well it's not so crazy". He got up and went back to Moscow. He had come all that way just to tell me that it wasn't so crazy.

ND So what did you do?

MGM  Nothing. But of course I told people about it. They changed their minds during that half year. But this was the scientific establishment including people very high. Then the war in South-East Asia was getting….

ND  But wasn't there a meeting between LBJ and Kosygin?

MGM  That's what I'm getting to. So things were a little dicky. But discussions went on and three years later Johnson and Kosygin talked at Glassborough. Macnamara presented our ideas. Kosygin listened and was not unpersuaded. Apparently there was a sort of understanding by that time and he must have talked to all these scientific types. That was three years later in I think 1967. And then I think there would have been a treaty. Or at least a draft treaty. But then the war in South-East Asia was getting hotter and in 1968 came the Prague spring and then the Russian troops marching in which made everything very dicky and so the whole issue was put off until finally in Nixon's administration they did something about it. I didn't play too much of a role in that except by that time I was in the President's Science Advisory Committee and of course I was in favour of such a treaty. But most of my activity was back in the first year.

ND  So you were essentially an initiator of those ideas?

MGM  No I wouldn't say that. But I was an early proponent. But there were a lot of other people thinking along those lines, including high officials in the Defense Department. But they also had other arguments in the opposite direction of course. I wouldn't take credit for initiating it. But I was one of the pioneers.

ND  But 1962, that's pretty early.

MGM  But in 1963 we had had another JASON study at Wood's Hole and there it was formally arranged for a group of us which included Pete Scoville who was a high official in the CIA and who later became an Arms Control bigshot

ND  I met him at a Pugwash[6] meeting

MGM Tom Schelling who was a Harvard economics professor, who was very high up in the strategy community and who was a very sceptical kind of person. He changed later and became much more human. And various others, all very distinguished. I chaired it for some reason although I was probably youngest. We then did a more formal job on it.

ND  Was there strong opposition?

MGM  Not within our group. Everyone agreed.

ND  But in other groups?

MGM  Oh well there was Teller and people like that of course. The same people as now. It hasn't changed. Exactly the same people.

ND  Has anything changed?

MGM  The issues have changed a little. Many more nations involved. The Soviet Union doesn't exist. Russia isn't really an enemy. There are these little tiny powers against which one might have to defend oneself. There's a whole bunch of different nuances. A lot of things have changed. But the basic fundamental disagreement hasn't changed.

ND  But forgetting tiny powers?

---

[6] Series of meetings on arms control between western and Soviet-bloc scientists.

MGM  But that's very important. That's the excuse for the system.

ND  I know it's an excuse.

MGM  It's very important. And there's also China

ND  But China's a major power.

MGM  But China didn't have anything then. But now it does and that's a huge issue today and that's another change in the situation. And the Pakistan and India situation is a huge change. And if we through the stupid system cause the Chinese to increase their offensive force, it's not only a nuisance for us but it will cause India to increase its forces, and then Pakistan. And that's supposed to be against our policy.

ND  So in some ways it is even more important now not to go down that route?

MGM  But we are not thinking of strikes between the US and the Soviet Union. That was the main worry then. The situation has changed. But the fundamental mind map has not changed.

ND  OK. Let's get on.

MGM  But I wanted to mention this group in 1963. You asked about classification. In 1963 we actually wrote a paper. We submitted it to the Defense Department. So people could read it. So it could influence policy. The problem was that I wanted very much to put the numbers in. To say "suppose we go all out and we build offensive weapons and they build offensive weapons and we build defensive weapons and they build defensive weapons. Where does this end?" What number of rockets would we have on each side? We decided the limiting factor ultimately was the number of nuclear weapons. So we included in the report these maximum numbers using the number of nuclear weapons which was an incredible secret. But that meant that nobody could read the report. Even the President couldn't read it. It was classified so high that nobody could read it.

ND  It's really funny

MGM  In the meantime in London a couple of young people associated with the IISS were trying to figure out what we were doing. If they had asked us we would have told them. Instead they were sitting there speculating.

ND  Presumably the report could be declassified taking out those numbers?

MGM  I don't know. I never saw the report. I'm not sure that I was allowed to read it.

ND  That's funny. In 1992 I was working on John Ward and wrote an article on 40 years of the H-Bomb. Ward came in and Sakharov. I published it. And asked Geoff West if Los Alamos Science would be interested in publishing it. It was a general article. Geoff showed it to the Editor at Los Alamos who was very keen. Then she was told that she couldn't publish it as the article was classified!

MGM  There's something else I should have mentioned about that summer study. In 1962 I had the advantage of talking with young Jasper Welch, an air force officer who was attached to this project. He was a young officer in the Air Force who used to think about strategy. He now lives in Santa Fe and I see him from time to time. He's retired of course and made it to major-general, maybe lieutenant-general. He and I talked a lot and he told me a lot of the ideas which were on the drawing board. So in my speeches, and I think also in the report, we put in stuff like MIRVs which didn't exist. Some people were quite impressed.

ND  Why did you put them in then?

MGM  Because we predicted them basically. We knew people were thinking about them. Because this would increase the instability. So we said that there are many forces making for instability and now we are going to add this stupid thing to it and we'll have a huge amount of instability. There are of course two different kinds: crisis instability and also instability in the arms competition. The MIRVs would be a serious problem here. So that was another feature. We were a little ahead of our time in the things we were willing to discuss. That may also have made the report difficult to access. Certainly MIRVs are no secret now.

ND  Let's change the subject. George Johnson talks about your political background. He mentions your socialist leanings when you were young.

MGM  Well they were more communist leanings. But it was only until I was 15 or so. The last spark of that kind of thing died with the Czech coup in February 1948 when I was 18. But I was disillusioned to a great extent long before that. But the last tiny trace of any sympathy with the Stalin outfit disappeared with that. Because Czechoslovakia was a place which could handle freedom and so on. They had known about it. It wasn't a place which had never known it. And to stamp it out there was terrible. The impression we had was of irreversibility. Before that coup there were parties and the Communist party was the biggest and so according to the constitution the President offered Gottwald the head of the Communist party the right to form a government and he formed one and then some of the socialists who were undercover communists like Zdenek Fuhrlinger switched sides and really revealed their true colours and then all of a sudden a government was formed legally and the next day it was all illegal; the next day the committees had taken over the factories and so on; the next day there wasn't freedom of speech any more; from one day to the next it had turned from a democracy with a strong communist party to some kind of dictatorship. That really impressed me.

ND  Even at 19.

MGM  No I wasn't 19 I was 18. But I had been thinking about all these thing since I was a little boy. I was very politically aware.

ND  But that was very precocious. Because lots of the left

MGM  But I was much more precocious than that. I was interested in these things when I was 5.

ND But by precocious I mean that I don't think 1948 made as big an impact on the left in Western Europe.

MGM But those people were impossible.

ND As 1956

MGM And even some of them lived through 1956 and continued to support the Soviet Union

ND Nevertheless during the early 1950s the McCarthyite period you would have considered yourself..

MGM No I had zero sympathy for these people after the Czech coup and I had very little before. For example in 1948 I supported Truman not Wallace. But towards the end of the war I still did. In 1944 when I was 15 and first went to Yale I still did.

ND Did you meet David Bohm?

MGM I met him. I described that in my book. That was in 1951 just before he left the country. I described it in connection with Quantum Mechanics.

ND Not in connection with politics?

MGM Well his problem was his Marxism. His Marxism led him to doubt quantum mechanics. That was an important part of my write-up. But I had no interest in Marxism.

ND Because there was a whole leftist fringe in the US

MGM What happened was during the 1930s before the Molotov-Ribbentrop Treaty in 1939 the Communists seemed to be the only people who were really militant about opposing Hitler and Mussolini. As well as Japanese aggression. And so many people who were horried by fascism and feared that it would conquer the world felt it was necessary to be friendly to the communists and adopt some of their ideas. That was very common and I think it was very understandable. After the Treaty though the whole thing became so obnoxious. So naturally I began to worry then in 1939. But when Hitler attacked the Soviet Union and the Soviet Union resisted so strongly I began again to feel that maybe they were the good guys. But by the end of the war I was not so interested any more and the final nail in the coffin was February 1948 when I lost any trace of interest. But it was already a few years since I was excited about it. But I don't know what George Johnson makes of it in his book.

ND Well he mentions Nina Byers

MGM Well that was different. That was what got me into trouble with my clearance. But by then I had no views of this kind. It was that I had lent my car to somebody who went to a Paul Robeson concert. The idea that that could get you into trouble. I was out of the country actually. No I told them that I was out of the country. It turned out that I had calculated the dates slightly wrong. I was in Princeton in the hospital and a couple of days later I left for Europe. So they probably looked at my passport records and concluded that I was lying as I said that I was in Europe. I was not in Chicago; I was in Princeton on my way to Europe that was what I should have said. But that was ridiculous. That was just absolutely ridiculous. Can you imagine all those FBI people wandering all around the neighbourhood taking down all the licence numbers.

ND  Yes

MGM  Another thing the authorities may not have liked was that back in 1943 a friend of my brothers who had some pro-Russian leanings got me a subscription to the Bulletin of the Soviet Embassy, which came to me for many years afterwards, until they asked for money for it and I no longer got it. But it came until for free from I think 1943 to 1947. And I used to read it. It was very interesting actually and I am very glad I read it. For example I read their version of the Katyn Forest massacre. As soon as the Soviets reached the site of that massacre they reported it as a Nazi atrocity. They made a huge thing about it. So I learned from this Bulletin about the Nazi atrocity. Later of course I learned that it was a Soviet atrocity. That was very interesting. If the authorities found out about my subscription which they probably did

ND  I'm sure they did

MGM  Then they wouldn't like that. But that ended in 1947. During that period from 44 to January 48 my sympathies declined very rapidly to zero.

ND  But you had some problems with your grant according to Johnson's book.

MGM  The Atomic Energy thing. It isn't that I had problems so much. It is that one communist named Hans Freischstadt got one of these AEC fellowships and some idiot Senators got very upset about it and they required every recipient of an Atomic Energy Commission Fellowship to have a full clearance. That was when I first encountered all these problems whatever they were. I never knew what they were.

ND  But you did get a full fellowship.

MGM  Oh Yes. That one came through. But it took me a year. For a year I had no money. I had to borrow from my friends. At the end of the year they gave me all the money in a lump sum. So I was able to pay them back and buy a car and so on.

ND  You weren't that friendly with Oppenheimer?

MGM  Very friendly.

ND  But that wasn't held against you?

MGM  No No I don't think that caused any problems

ND  But Frank Oppenheimer

MGM  No the other way round. Frank Oppenheimer was a communist. So there wasn't a problem that he knew Robert.

ND  But Frank was one of the reasons that Robert was under suspicion.

MGM  That's right. But I don't know anyone who got into trouble by knowing Robert. There was just a conspiracy against him by those right wing kooks. That was terribly unfortunate I was so angry about that. Really really really angry.

ND  Do you speak to Teller?

MGM  I did once or twice. For a long time I wouldn't and now I wouldn't again. When I moved here I got a message saying he would like to see me. I sent a message back saying I wouldn't like to see him. The only time we ever spoke after that was at a meeting of a committee I was on to decide about some classification questions. Teller wanted everything declassified.

ND  Why did he want everything declassified?

MGM  I suppose he felt that he would get greater glory if it was declassified. In recent times he has been in favour of declassification. We thought that it was dangerous for people in various countries to be able to make progress with their development of weapons, including thermonuclear weapons, if all these things became public and he basically didn't care as he thought some of these countries would be South Africa and Israel and so on which he thought ought to have hydrogen bombs.

ND  Did you see the New York Times piece a couple of weeks ago which said that it was Dick Garwin rather than Teller who was mainly responsible for the first H-Bomb?

MGM  Dick Garwin?  Not Stan Ulam?

ND  I don't know Ulam's contribution…

MGM  He wrote a paper on it.  Teller wrote it with him. I've seen the paper. In the paper it seems fairly clear that there were two strands, two different styles, basically two different subjects and the interesting stuff was Ulam's and talking on and on about the Super was obviously Teller's contribution. So although they are meant to share credit, in my mind Ulam's contribution was more significant.

ND  Teller had a heart attack about 20 years ago and when he was recovering he dictated some notes about all that and they surfaced in the New York Times a couple of weeks ago. Teller didn't give Ulam much credit.

MGM  He certainly didn't.

ND  He said that Garwin was the guy who got the design together of , which was it, that enormous bomb, the very big one

MGM  When in 1952. The Eniwetok test.

ND  Yes. Dick Garwin was responsible for seeing the thing through.

MGM  But that wasn't a real hydrogen bomb. It was the kind of thing you had to take to the target by

ND  Sure. But it was the first H- bomb which achieved 20 Megatons or whatever.

MGM  It was a big blast but it wasn't useable. It was not a real weapon.

ND  Nevertheless it illustrated the principles.

MGM Dick of course is a very smart fellow.

ND  I'm exhausted.

MGM  Anyway if you include a few things about simplicity and complexity and conservation it will be interesting. I call it plectics. Plectics and nature conservation. It doesn't have to be a lot.

ND  Quantum Cosmology?

MGM  Well I haven't made any real contributions to quantum cosmology. What I have made contributions to along with Jim is to quantum mechanics as it has to be since we have quantum cosmology. The Copenhagen interpretation isn't much use to quantum cosmology. So we worked on how to do quantum mechanics properly. God know what George said about that if anything. He couldn't possibly have got it right.

ND  I'll have to read your papers

MGM  Well you can look at my book. It clearly says most of that.

ND  This one?

MGM  Yes. You haven't read it?

ND  I've skimmed through it.

MGM It's worth reading for your project.

OK I'll read it.

MGM  What about all the reminiscence papers. You've got those.

ND  Yes Olivia[7] sent them to me. But I needed a framework in my own mind first.

---

[7] MGM's secretary